Optimal merging from an on-ramp into a high-speed lane dedicated to connected autonomous vehicles

L.C. Davis

10244 Normandy Dr., Plymouth, MI 48170, United States

ABSTRACT

In the future, High Occupancy Vehicle (HOV) lanes or other dedicated lanes might be restricted to autonomous vehicles, *e.g.* wirelessly connected vehicles with longitudinal motion control. These vehicles would likely travel at high speeds in platoons. New criteria for the merging of a vehicle from an on-ramp are proposed. To reduce disruption to the flow, only merges into gaps between platoons, not within, are considered. To minimize HOV lane trip time, vehicle acceleration and deceleration, the optimal merge position is determined from simulations of linear combinations of the deviation of the headway from equilibrium and vehicle velocity differences. These are between the merging vehicle and the lead vehicle (on the HOV lane) and between the trailing vehicle (on the HOV lane) and the merging vehicle. The merging vehicle, due to acceleration limitations on the on-ramp, generally will merge at a significantly lower velocity than the HOV lane average velocity. A queue of vehicles is held at the entrance to the on-ramp waiting for a suitable gap between platoons to approach.

**Highlights**

- Freeway lane dedicated entirely to wirelessly connected autonomous vehicles
- Vehicles have the capabilities of cooperative adaptive cruise control vehicles
- Vehicles travel at high speed in platoons
- New criteria for merging from an on-ramp into gaps between platoons

1. Introduction

There have been several proposals to devote High Occupancy Vehicle (HOV) lanes or other lanes dedicated exclusively to autonomous vehicles [1-3]. Presumably the velocity of such vehicles in the HOV lane would be substantially higher than velocities in normal freeway traffic. When the HOV lane is strictly dedicated to autonomous vehicles, only other autonomous vehicles can merge. For simplicity, I use the designation "HOV" to apply to any such dedicated lane.

This paper addresses merging of vehicles that are equipped with vehicle-to-vehicle communication and a longitudinal control system (adaptive cruise control or a more elaborate connected system), known as Cooperative Adaptive Cruise Control (CACC) vehicles [4]. It is expected that vehicles in the HOV lane travel in platoons of various numbers of vehicles. Within the platoon, control systems attempt to keep vehicles at the equilibrium gap, $hv$, where $h$ is the headway time (~1 s) and $v$ is the average platoon velocity.

To minimize disruption of a platoon, in this paper only the regions between platoons are considered suitable for merging. Thus, a merging vehicle might join a platoon at its end, but insertion into the interior of a platoon is not allowed. In this respect, it apparently differs from the algorithm developed at the Distributed Intelligent Systems and Algorithms Laboratory at the Ecole Polytechnique Federale de Lausanne (EPFL) as described by Sabin [5]. (To my knowledge, the actual algorithm has not been published.) The emphasis in the present work is to minimize the acceleration, deceleration and trip time delay of vehicles in the HOV lane caused by merging vehicles.

This situation differs from merging from an on-ramp into a normal freeway lane where the merging vehicle can generally accelerate to the prevailing freeway speed [6]. Even if a suitable gap between vehicles exists in a high-speed HOV lane, the trailing vehicle might be required to decelerate for the merge to occur because the merging vehicle cannot be expected to accelerate enough on an on-ramp. Thus, the position of the merge within the gap must be chosen carefully.

The literature on merging and lane changes is extensive, although no paper has addressed entering a HOV lane devoted entirely to autonomous vehicles. Treiber and Kesting [7] discuss the MOBIL algorithm (which stands for "minimizing overall braking deceleration induced by lane changes"). The principal criterion is that the deceleration of the following vehicle (the one in the receiving lane just behind the merged vehicle) be no more than a safe value (2 ms$^{-2}$). They also include a bias term and a politeness factor to determine the acceleration of the vehicle making the lane change. Rios-Torres and Malikopoulous [8] consider optimizing the control input to minimize fuel consumption. Ntousakis, Nikolos and Papageorgiou [9] describe longitudinal trajectory planning that minimizes not only acceleration, but first and second derivatives, of the merging vehicles. Wang, Wu and Barth [10] propose a distributed consensus algorithm for gap creation based on V2V communication. Vehicle control is by "ghost" vehicles to which the merging and following vehicles respond. Scarini, Hegyi and Heydecker [11] define a merging assistant strategy that relies on V2V communication and Vehicle-to-Infrastructure (V2I) communication. The Cooperative Merging Assistant (CoopMA) creates platoons and gaps (for merging vehicles) through cooperative vehicles that slow down in a controlled manner on the freeway [12]. Advanced Driver Assistant Systems (ADAS) consisting of vehicles with ACC and automated emergency braking are considered by Altche', Qian and de La Fortelle [13] who propose a supervised coordination scheme guaranteeing safety and deadlock avoidance to override driver commands where necessary. Katrakazas, Quddus, Chen and Deka [14] review real-time motion planning methods for merging, encountering intersections and obstacle avoidance. Morales and Nijmeijer [15] evaluate a cooperative tracking controller to keep a certain distance between vehicles. Nishi, Doshi, James and Prokhovov [16] apply a multipolicy decision learning method called passive-actor critic to freeway merging. Using neural networks, Wang and Chan [17] also apply reinforcement learning to merging. Each of these papers provides further references to earlier research on merging. For further information about the effects of cooperative adaptive cruise control and vehicle communications no traffic flow and stability, see Refs. [18-24].

The organization of this paper is Sec. 2 describes the model, Sec. 3 presents simulations, and Sec. 4 contains conclusions.

2. Model

In the present analysis the longitudinal control algorithm for the desired acceleration of an ACC vehicle is [25-27]

$$a_n^d = \frac{\alpha}{h}(x_{n-1} - x_n - D - hv_n) + k(v_{n-1} - v_n) - \xi a_n. \tag{1}$$

The vehicle in front of the vehicle of interest (designated $n$) is the "lead" and the parameters are the sensitivity $\alpha$, the headway time constant $h$, the coefficient of relative-velocity feedback $k$, and the acceleration-feedback gain $\xi$. $D$ is the vehicle length (plus a safety margin). With a mechanical response time, the vehicle dynamics are then given by

$$\tau \frac{da_n}{dt} + a_n = a_n^d, \tag{2}$$

with the limits imposed on acceleration and velocity such that

$$-d_{max} \leq a_n^d \leq a_{max}, \tag{3a}$$

$$0 \leq v_n \leq v_{max}. \tag{3b}$$

The maximum acceleration, deceleration and velocity are $a_{max}$, $d_{max}$ and $v_{max}$. For simplicity, all vehicles are taken to be identical.

The vehicle attempting to merge is designated $m$ and the gap is between vehicles $a$ and $b$. See Fig. 1a. The minimum size gap considered is $2hv_{max} + D$. Thus $x_b - x_a = 2(hv_{max} + D)$. In the proposed method, two conditions must be satisfied for the merge to happen:

$$S_a = x_a - x_m - D - hv_m + T_v(v_a - v_m) \geq 0, \tag{4a}$$

$$S_b = x_m - x_b - D - hv_b + T_v(v_m - v_b) \geq 0, \tag{4b}$$

where, to obtain optimal results, the parameter $T_v$ is chosen by numerical simulation.

As illustrated in Fig. 1b, the merge point is at least a distance $T_v(v_{max} - v_m)$ before the midpoint when the conditions of Eq. (4) are satisfied.

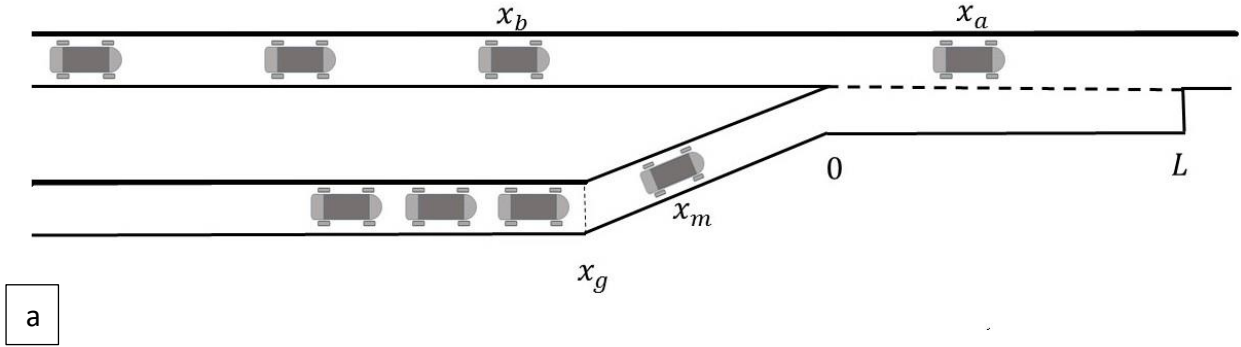

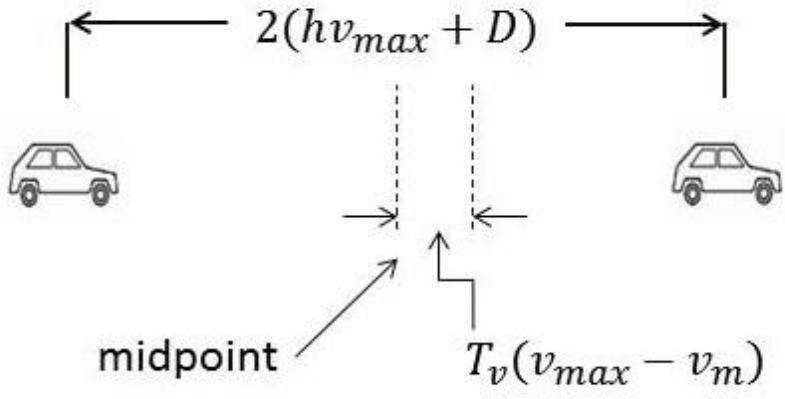

Fig. 1. (a) On-ramp to high-speed HOV lane. Vehicle $m$ waits at $x_g$ for a suitable gap between vehicles $a$ and $b$ travelling at $v_{max}$ to arrive. When vehicle $m$ is released, it accelerates to velocity $v_m$ by the time it reaches the entrance ($x = 0$) to the merge region of length $L$. A queue of vehicles (for eventual merging) exists to the left of $x_g$. (b) The merge point of the smallest gap considered is at least $T_v(v_{max} - v_m)$ before the midpoint.

After a merge, the desired acceleration of vehicle $m$ is

$$a_m^d = \max\{-d_{max}, \min\{a_{max}, \frac{\alpha}{h}(x_a - x_m - D - hv_m) + k(v_a - v_m) - \xi a_m\}\}. \tag{5}$$

In some instances, if $v_b - v_m$ is large enough, additional braking at a rate higher than $d_{max}$ can be required to avoid a collision. See Appendix A.

The measure of acceleration of vehicles in the HOV lane is

$$a_{tot} = \left[\frac{1}{MT_{max}}\Sigma_n \int_0^{T_{max}} dt\, \theta(\ddot{x}_n)(\ddot{x}_n)^2\right]^{1/2}, \tag{6}$$

where the sum is over all vehicles in the HOV lane and $M$ is the number of merges in time $T_{max}$. $\theta(\ )$ is the Heaviside function (1 for positive argument; 0, otherwise). Normalizing to $T_{max}$ is essentially normalizing to the number of vehicles passing the merge region because the flow of main line vehicles is constant. The measure of deceleration is

$$d_{tot} = \left[\frac{1}{MT_{max}}\Sigma_n \int_0^{T_{max}} dt\, \theta(-\ddot{x}_n)(\ddot{x}_n)^2\right]^{1/2}, \tag{7}$$

The delay in trip time is the actual time to travel from origin to destination minus the time if the vehicle were to travel at $v_{max}$. The average over all vehicles that originated upstream of the merge region (does not include merged vehicles) is denoted by $t_{ave}$.

The region that a vehicle is permitted to enter the HOV lane (dashed line in Fig. 1) is the zone $0 < x < L$. Only one vehicle can merge at a time. The condition $S_b > 0$ and $S_a > 0$ must be satisfied for the merge to occur. In addition, a minimum gap of 10 m between vehicles $a$ and $m$ is required.

Vehicle $m$ waits to merge at $x_g$ for a suitable gap on the HOV lane to approach. For any pair $a$ and $b$ of consecutive vehicles for which

$$x_b < 0, \tag{8}$$

and

$$x_a \geq x_b + 2(hv_b + D), \tag{9}$$

the following arrival times at the entrance to the merge region $(0 < x < L)$ are estimated (once every 0.1 s in the simulations)

$$T_a = -\frac{x_a}{v_a}, \tag{10a}$$

$$T_b = -\frac{x_b}{v_b}, \tag{10b}$$

$$T_m = \sqrt{\frac{-2x_g}{a_{max}}}. \tag{10c}$$

If $T_a < T_m < T_b$ the vehicles would arrive at the entrance in the correct sequence. Furthermore, if the following holds, vehicle $m$ can be released from $x_g$ when:

$$T_m > T_a + \frac{D}{v_a} + (h + T_v)\frac{v_{m0}}{v_a} - T_v, \tag{11a}$$

$$T_m < T_b - \frac{D}{v_b} - h - T_v + T_v\frac{v_{m0}}{v_b}, \tag{11b}$$

where $v_{m0} = a_{max}T_m$ is the expected velocity of vehicle $m$ when $x_m = 0$. These conditions place vehicle $m$ in approximately the correct position to merge as it enters the merge region.

When $x_g < x_m < 0$ (after release) the desired acceleration is

$$a_m^d = \min\{k(v_{m0} - v_m), a_{max}\}. \tag{12}$$

Once $x_m > 0$ and vehicles a and b verify (system checks every 0.1 s) that the gap between them is large enough (gap $\geq 2hv_{max} + D$) and $x_b < x_m < x_a$ then Eq. (5) applies even though vehicle m has not yet merged. Additionally, if $S_b < 0$, vehicle b is required to decelerate at

$$a_b^d = -d_{max}. \tag{13}$$

If verification fails, then vehicle m must accelerate (or decelerate) with

$$a_m^d = \min\{\max[-d_{max}, A_m - \xi a_m], a_{max}\}, \tag{14a}$$

where

$$A_m = \frac{\alpha}{h}(x_a - x_m - hv_m) + k(v_a - v_m), \tag{14b}$$

or

$$A_m = -\left[\frac{\alpha}{h}(x_m - x_b - hv_b) + k(v_m - v_b)\right]. \tag{14c}$$

The former applies if $S_a < 0$ and $S_b > 0$ and the latter if $S_a > 0$ and $S_b < 0$. [Note that $S_a + S_b > 0$. See Eq. (4).]

If no merge occurs before $x_m$ reaches the midpoint ($L/2$), then if $S_a < 0$ the desired acceleration of vehicle $m$ is

$$a_m^d = -\frac{d_{max}}{2}. \tag{15}$$

If $S_b < 0$ then $a_m^d = 0$ and the desired acceleration of vehicle $b$ must be

$$a_b^d = -d_{max}. \tag{16}$$

Under some conditions, vehicle $b$ must decelerate at a larger rate $\frac{3}{2}d_{max}$ following a merge. See Appendix A.

V2V communication between vehicle m and vehicles a and b is required to establish a suitable gap for merging and when vehicle m should be released from $x_g$. Also, communication between vehicles m and b is needed when b needs to brake. It is assumed that the signals travel with delay or packet losses. For this ideal system, where the necessary computations and decisions take place is not specified.

It is reasonable to assume that the vehicles on the HOV lane would travel in platoons. In simulations the number of vehicles in a platoon is taken to be $N_{gap} + 1$ where the number of gaps is

$$N_{gap} = \max\{2, Int(1 + Rnd\ N_{plat})\}, \tag{17}$$

where $Rnd$ is a random number between 0 and 1 and $Int$ is the integer part. Thus, $N_{gap} = 2, 3 \ldots N_{plat}$. The initial gap between vehicles in a platoon is $hv_{max}$. The separation (front bumper to rear bumper plus $D$) of platoons is given by (See Fig. 2.)

$$L_{sep} = \max\{1, Rnd\, L_{plat}\}(hv_{max} + D). \tag{18}$$

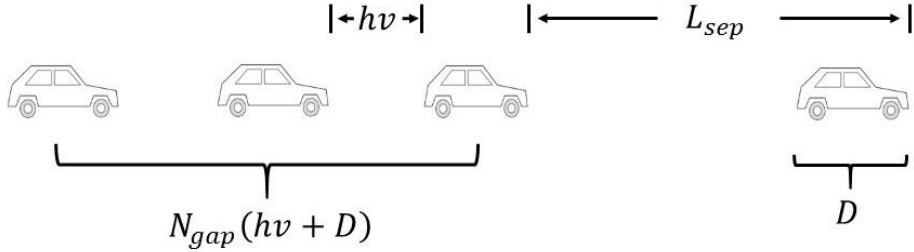

Fig.2 Schematic of a platoon made up of $N_{gap} + 1$ vehicles separated by $L_{sep}$ from the last vehicle of the preceding platoon.

The average incoming free flow rate is $\frac{(\langle N_{gap}\rangle+1)v_{max}}{\langle L_{sep}\rangle+\langle N_{gap}\rangle((hv_{max}+D)}$ where

$$\langle N_{gap}\rangle = \frac{N_{plat}+1}{2} + \frac{1}{N_{plat}}, \tag{19a}$$

and

$$\langle L_{sep}\rangle = \left(\frac{L_{plat}^2-1}{2\,L_{plat}} + \frac{1}{L_{plat}}\right)(hv_{max} + D). \tag{19b}$$

For $L_{plat} = 5$ and $N_{plat} = 6$, the average flow of 2239 vehicles/h, a substantial fraction of the maximum of 3007 vehicles/h, when $v_{max} = 38$ m/s.

3. Simulations

Simulations of this model are presented in this section. The values of parameters are given in Table 1.

Table 1. Parameters

| | |
|---|---|
| $\alpha$ | 2 s$^{-1}$ |
| $h$ | 1 s |
| $k$ | 1 s$^{-1}$ |
| $\tau$ | 0.5 s |
| $D$ | 7.5 m |
| $\xi$ | 0.6 |
| $d_{max}$ | 2 ms$^{-2}$ |
| $a_{max}$ | 3 ms$^{-2}$ |
| $L$ | 500 m |
| $|x_g|$ | 150 m |

The flow coming into the merge region on the HOV lane in Figs. 3 and 4 is determined by $L_{plat}$ = 5 and $N_{plat} = 6$, which gives on average 0.61 vehicles/s at $v_{max} = 38$ m/s. The merging vehicles enter the merge region at $v_m = 28$ m/s. The first vehicle in a queue of vehicles at $x_g$ waits on average less than 20 s to merge. Data are calculated from averages of twenty-five runs of 20,000 s each. Fig. 3 shows the dependence on $T_v$. The delay per vehicle on the HOV lane due to the merging vehicles, $t_{ave}$, reaches a broad minimum at about 0.01 s for $T_v = 2.5$ s compared to almost 0.08 s if $T_v = 0$ (Fig. 3a). Likewise, the measures of acceleration and deceleration are minimized at the same value of $T_v = 2.5$ s (Fig. 3b). The rate of merging (Fig. 3c) is the largest at a slightly lower value $T_v = 2.0$ s.

The contribution of the acceleration of the merged vehicle (which accelerates to $v_{max}$ from $v_m$) to $a_{tot}$ is approximately $\sqrt{a_{max}(v_{max} - v_m)/T_{max}}$. For $v_{max}$= 38 m/s, $v_m$= 28 m/s and $T_{max}$=2×10⁴ s, this amounts to 0.039 m/s², which is approximately the simulation value. On the other hand, $d_{tot}$ is primarily determined by the deceleration of trailing vehicles (labelled $b$ in Fig. 1).

A comparison of the values of $t_{ave}$, $a_{tot}$ and $d_{tot}$ for $T_v = 0$ and 2.5 s is shown in Fig. 4 as a function of HOV maximum velocity $v_{max} = 33$ to 38 m/s. The substantial reduction (at each $v_{max}$) of $t_{ave}$ and $d_{tot}$ when $T_v = 2.5$ s is evident. The measure of acceleration is less sensitive to $T_v$.

In Fig. 5, results for $L_{plat}$= 10 and $N_{plat}$= 2 are shown. In this rather different scenario; the platoons are small (just pairs of vehicles) and the separation of platoons from one another is large. Yet, the same dependence on $T_v$ is found.

Calculations (not shown) for other values of $v_m$ show the same trends as in Figs. 3-5.

4. Conclusions

The primary conclusion that can be drawn from the simulations presented in this paper is that whenever a vehicle enters a lane devoted to high-speed autonomous vehicles travelling in platoons, the following rules provide an optimal merging method. Merges are restricted to gaps between platoons and the rules are based on two linear combinations of the spatial gap compared to the equilibrium gap and the difference in velocities, one for the lead vehicle and the merging vehicle ($S_a$), and another for the merging vehicle and the trailing vehicle ($S_b$). Simulations indicate that the coefficient of the velocity difference $T_v$ should be 2.5 s to reduce the delay in travel time and to minimize acceleration and deceleration, without significantly reducing the rate of merging.

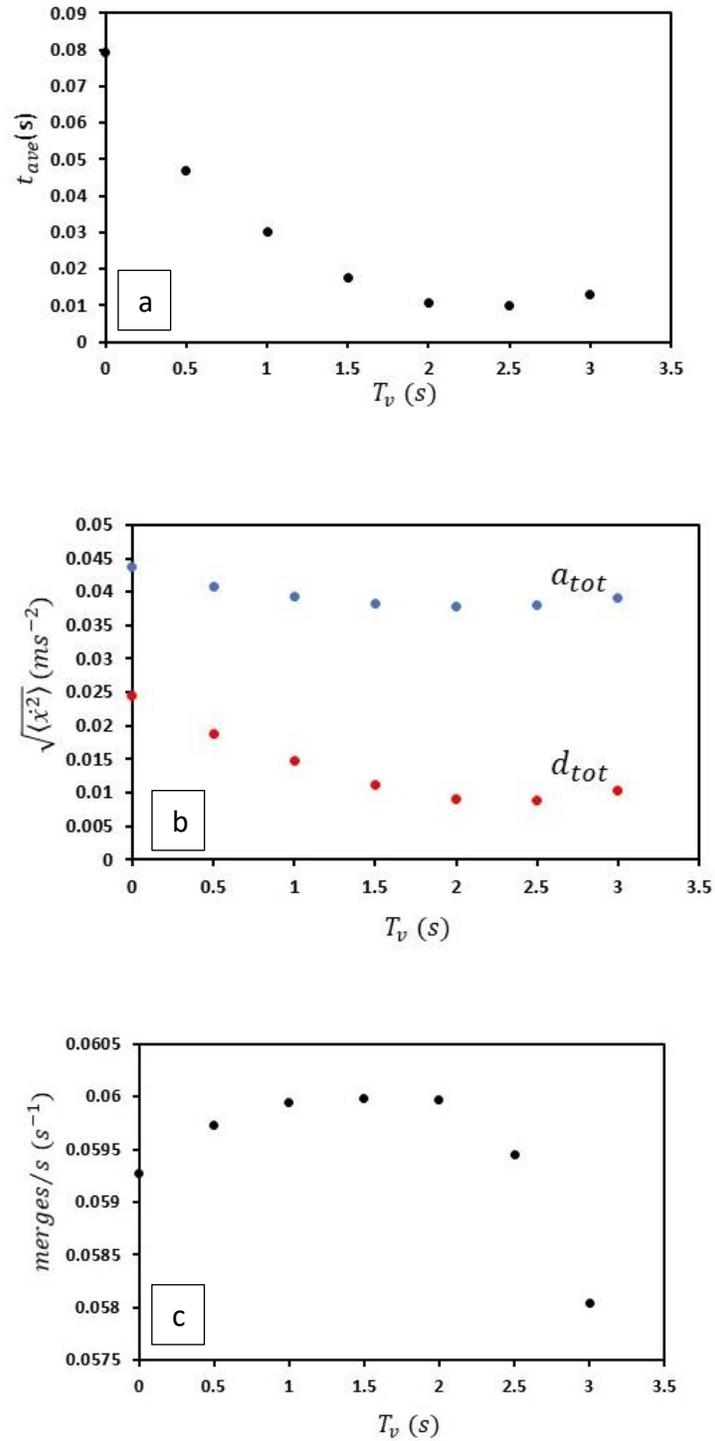

Fig. 3. Simulation results as a function of $T_v$ for $L_{plat}$ = 5 and $N_{plat} = 6$, corresponding to incoming flow of 0.61 vehicles/s. The velocity of merging vehicles is approximately 28 m/s and main line (HOV) vehicles travel at 38 m/s. (a) The delay in trip time of HOV

vehicles due to merging vehicles. (b) Measures of the acceleration $a_{tot}$ (blue) and deceleration $d_{tot}$ (red). (c) The rate of merging into gaps between platoons.

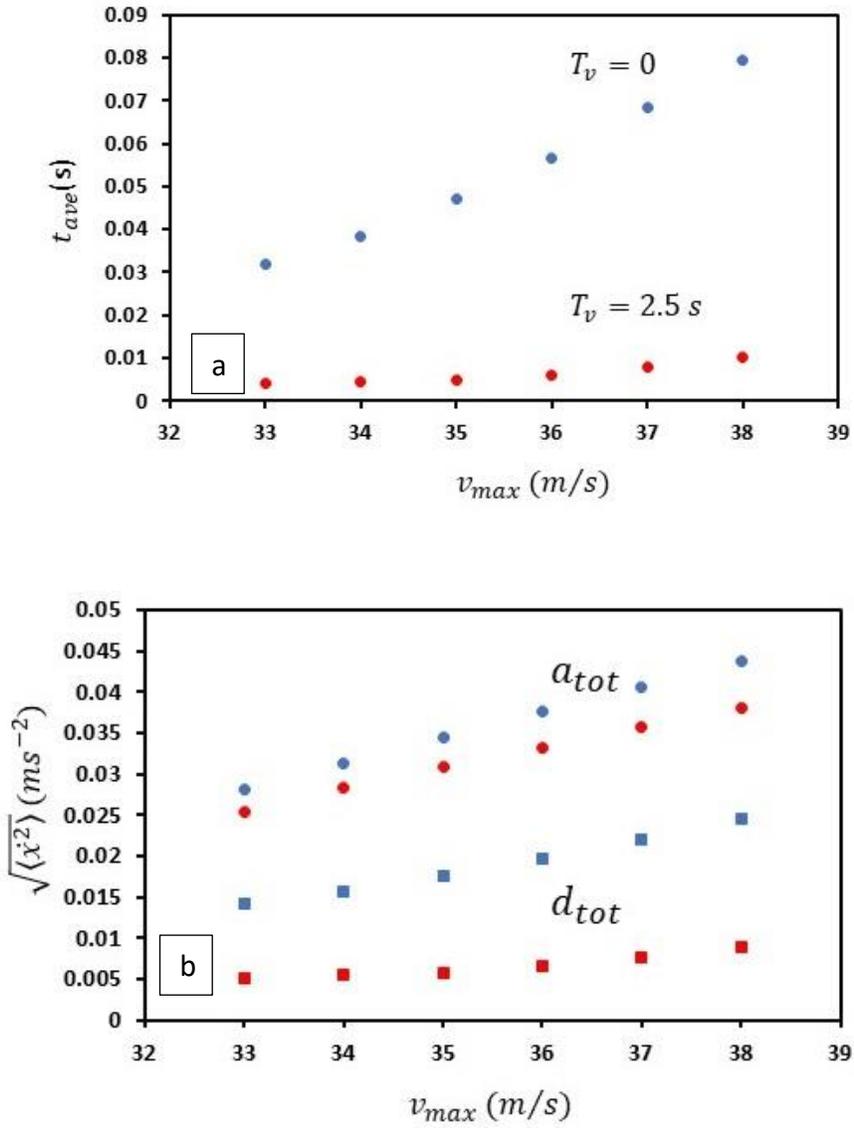

Fig 4. The delay $t_{ave}$ (a) and measures of acceleration and deceleration (b) as a function of the HOV lane maximum velocity $v_{max}$. Merging vehicles enter at approximately 28 m/s. For each quantity, values for $T_v$ =0 (blue) and 2.5 s (red) are shown. Other parameters the same as in Fig. 3.

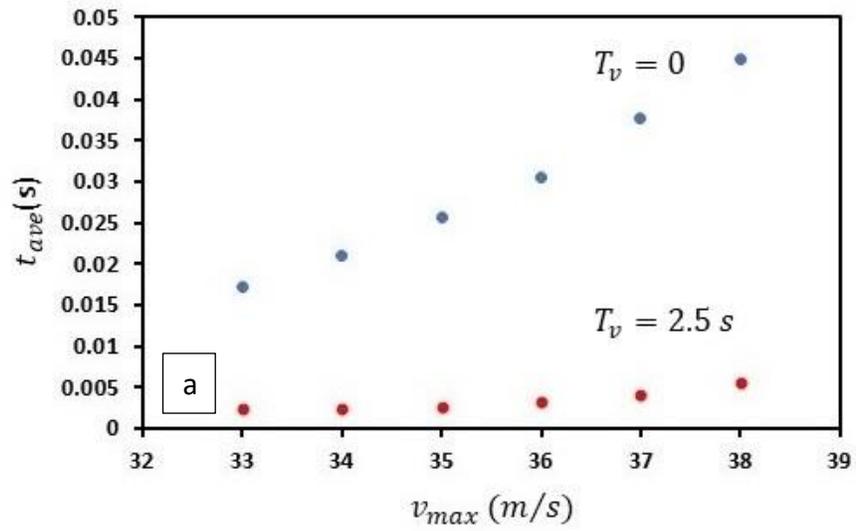

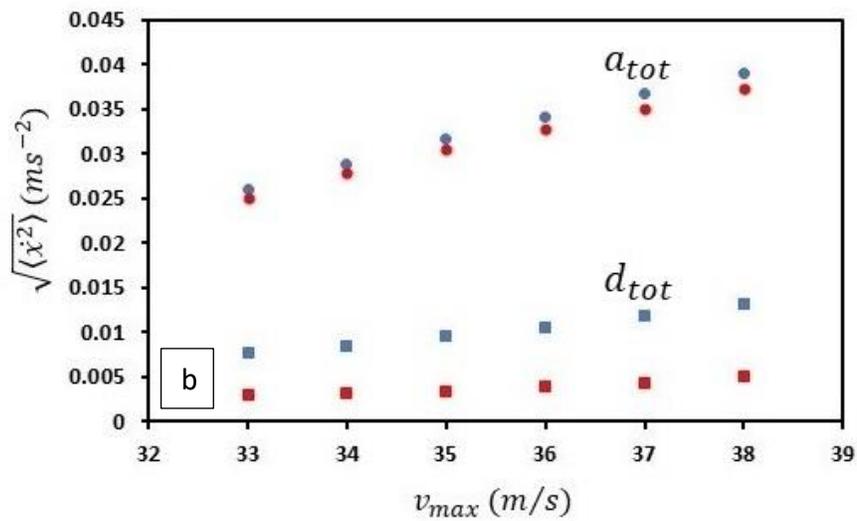

Fig. 5. Results for $L_{plat}$= 10 and $N_{plat}$= 2 as a function of $v_{max}$ with $v_m$ =28 m/s. (a), $t_{ave}$ for $T_v$ =0 (blue) and 2.5 s (red). (b), $a_{tot}$ (upper data) and $d_{tot}$ (lower data) for $T_v$ =0 (blue) and 2.5 s (red).

Appendix A. Additional deceleration

If the velocity difference $v_b - v_m$ at merge is large enough, the subsequent deceleration of $b$ can exceed an acceptable level. If $\dot{v}_m = 0$, it can be shown that the maximum deceleration is (neglecting response time) is

$$\max decel = \frac{\lambda_1 \lambda_2}{\lambda_1 - \lambda_2}(v_{max} - v_m)(e^{\lambda_1 \theta} - e^{\lambda_2 \theta}), \tag{A1}$$

where

$$\theta = \frac{\ln(\frac{\lambda_2}{\lambda_1})}{\lambda_1 - \lambda_2}, \tag{A2}$$

and

$$\lambda_{1,2} = \frac{1}{2}\left[-(\alpha + k) \pm \sqrt{(\alpha + k)^2 - 4\frac{\alpha}{h}}\right]. \tag{A4}$$

To avoid reaching maximum deceleration set $a_b^d = -d'_{max}$ as soon as $x_b = (x_m - D - hv_b) + \left(\frac{hk}{\alpha}\right)(v_m - v_b) < 0$ [note $v_b = v_{max}$ at this time] and continue until $\frac{\alpha}{h}(x_m - x_b - D - hv_b) + k(v_m - v_b) > -d'_{max}$ and $v_m < v_b < v_m + d'_{max}T$ where

$$T = [\frac{\lambda_1 \lambda_2}{\lambda_1 - \lambda_2}(e^{\lambda_1 \theta} - e^{\lambda_2 \theta})]^{-1}. \tag{A5}$$

Taking $d'_{max} = 1.5\, d_{max}$ was found to be adequate for the simulations of this paper.

References

[1] Jack Stewart, Transportation 10.08.1708:00 am, Maybe it's time to cede us freeways to driverless cars, https://www.wired.com/story/self-driving-cars-take-over-highways/

[2] Lanhang Ye, Toshiyuki Yamamoto, Impact of dedicated lanes for connected and autonomous vehicle on traffic flow throughput, Physica A 512 (2018) 588–597.

[3] Gary Richards, Roadshow: Reserve lanes for self-driving cars to ease traffic congestion, https://www.mercurynews.com/2018/03/05/roadshow-reserve-lanes-for-self-driving-cars-to-ease-traffic-congestion/

[4] Steven E. Shladover, Christopher Nowakowski, Xiao-Yun Lu, Robert Ferlis, Cooperative adaptive cruise control (CACC) definitions and operating concepts 2015, 4th TRB Annual Conference, Washington D. C Transp. Res. Record J. Transp. Res. Board TRB, Vol.2489, (2015) 145–152.

[5] Dyani Sabin, New Algorithm Lets Self-Driving Cars Merge With Traffic Like a School of Fish, on January 30, 2017, www.inverse.com/article/27119-algorithm-merge-autonomous-highway; also phys.org/news/2017-01-driver-vehicles-cooperate.html.

[6] Boris S. Kerner, Breakdown in Traffic Networks, Springer, Berlin, 2017. See Sec. A.6.3, p.568.

[7] Martin Treiber, Arne Kesting, Traffic Flow Dynamics Data, Models and Simulation, Springer Heidelberg, 2013.

[8] Jackeline Rios-Torres, Andreas A. Malikopoulos, automated and cooperative vehicle merging at highway on-ramps, IEEE Transactions on Intelligent Transportation Systems 18 (2017) 780-9.

[9] Ioannis A. Ntousakis, Ioannis K. Nikolos, Markos Papageorgiou, Optimal vehicle trajectory planning in the context of cooperative merging on highways, Transportation Research Part C 71 (2016) 464–488.

[10] Ziran Wang, Guoyuan Wu, Matthew J. Barth, Developing a distributed consensus-based cooperative adaptive cruise control system for heterogeneous vehicles with predecessor following topology, Journal of Advanced Transportation (2017) Article ID 1023654, 1- 16.

[11] Riccardo Scarinci, Benjamin Heydecker, Control Concepts for Facilitating Motorway On-ramp Merging Using Intelligent Vehicles, Transport Reviews, 34 (2014) 775–797.

[12] Riccardo Scarinci, Benjamin Heydecker, Andreas Hegyi, Analysis of traffic performance of a merging assistant strategy using cooperative vehicles, IEEEITSC Article Special Issues_2014.

[13] Florent Altche', Xiangjun Qian, Arnaud de La Fortelle, An algorithm for supervised driving of cooperative semi-automous vehicles (extended), arXiv:1706.008046v1 [cs.MA] 25 Jun 2017.

[14] Christos Katrakazas, Mohammed Quddus, Wen-Hua Chen, Lipika Deka, Real-time motion planning methods for autonomous on-road driving: State-of-the-art and future research directions, Transportation Research Part C 60 (2015) 416–442.

[15] América Morales, Henk Nijmeijer, Merging strategy for vehicles by applying cooperative tracking control, IEEE Transactions on Intelligent Transportation Systems 17 (2016) 3423-33.
13

Figure Captions

Fig. 1. (a) On-ramp to high-speed HOV lane. Vehicle $m$ waits at $x_g$ for a suitable gap between vehicles $a$ and $b$ travelling at $v_{max}$ to arrive. When vehicle $m$ is released, it accelerates to velocity $v_m$ by the time it reaches the entrance ($x = 0$) to the merge region of length $L$. A queue of vehicles (for eventual merging) exists to the left of $x_g$. (b) The merge point of the smallest gap considered is at least $T_v(v_{max} - v_m)$ before the midpoint.

Fig.2 Schematic of a platoon made up of $N_{gap} + 1$ vehicles separated by $L_{sep}$ from the last vehicle of the preceding platoon.

Fig. 3. Simulation results as a function of $T_v$ for $L_{plat}$ = 5 and $N_{plat} = 6$, corresponding to incoming flow of 0.61 vehicles/s. The velocity of merging vehicles is approximately 28 m/s and main line (HOV) vehicles travel at 38 m/s. (a) The delay in trip time of HOV vehicles due to merging vehicles. (b) Measures of the acceleration $a_{tot}$ (blue) and deceleration $d_{tot}$ (red). (c) The rate of merging into gaps between platoons.

Fig 4. The delay $t_{ave}$ (a) and measures of acceleration and deceleration (b) as a function of the HOV lane maximum velocity $v_{max}$. Merging vehicles enter at approximately 28 m/s. For each quantity, values for $T_v$ =0 (blue) and 2.5 s (red) are shown. Other parameters the same as in Fig. 3.

Fig. 5. Results for $L_{plat}$= 10 and $N_{plat}$= 2 as a function of $v_{max}$ with $v_m$ =28 m/s. (a), $t_{ave}$ for $T_v$ =0 (blue) and 2.5 s (red). (b), $a_{tot}$ (upper data) and $d_{tot}$ (lower data) for $T_v$ =0 (blue) and 2.5 s (red).